\documentclass [aps,pra,twocolumn,amsfonts,graphics]{revtex4}
\usepackage{graphicx}
\newcommand {\beq}{\begin{equation}}
\newcommand {\eeq}{\end{equation}}
\newcommand {\beqa}{\begin{eqnarray}}
\newcommand {\eeqa}{\end{eqnarray}}
\newcommand {\la} {\langle}
\newcommand {\ra} {\rangle}

\begin{document}
\title{Cloning quantum entanglement in arbitrary dimensions}
\author{E.~Karpov, P.~Navez, and N.~J.~Cerf}
\address{Quantum Information and Communication, \'Ecole Polytechnique,
CP 165, Universit\'e Libre de Bruxelles, 1050 Brussels, Belgium}

\begin {abstract}
We have found a quantum cloning machine that optimally duplicates 
the entanglement of a pair of $d$-dimensional quantum systems. 
It maximizes the entanglement of formation contained in the two copies 
of any maximally-entangled input state, while preserving 
the separability of unentangled input states.  Moreover, it cannot increase
the entanglement of formation of all isotropic states.
For large $d$, the entanglement of formation of each clone tends
to one half the entanglement of the input state, which corresponds to
a classical behavior. Finally, we investigate a local entanglement cloner,
which yields entangled clones with one fourth the input entanglement
in the large-$d$ limit.
\end {abstract}


\maketitle
\section {Introduction}

The no-cloning theorem \cite{noclo} precludes a perfect cloning of arbitrary quantum states. However, an imperfect cloning is possible and various quantum cloning machines (QCM), which duplicate quantum states 
with the highest fidelity, have been proposed following the seminal paper
of Buzek and Hillery \cite{QCM}.
Recently, the question of whether quantum entanglement can be cloned 
or not was raised in \cite{LNFC04}. Since the quantum entanglement 
is a resource for quantum computation, quantum communication, and quantum cryptography, it is important to know up to what extent this resource can be duplicated. For maximally-entangled (ME) states in two dimensions, an
{\it entanglement no-cloning principle} was formulated :
``if a quantum operation can be found that perfectly duplicates the entanglement
of all ME states, then it is necessary does not preserve separability''.
A QCM was proposed that optimally (but imperfectly) clones the entanglement
of two-dimensional bipartite states (qubits) while preserving separability.

In the present paper, we extend these results
to pairs of $d$-dimensional quantum systems, with arbitrary $d$. 
We show that a (symmetric) cloning machine
can be defined, which maximizes the amount of entanglement of the two clones
of ME states, while producing separable clones in the case of unentangled
input states. We analyze the entanglement of the clones in terms 
of fidelity, but show that optimizing the cloning machine in terms 
of fidelity actually leads to maximizing the entanglement of formation
of the clones provided that we restrict to cloning machines 
that are covariant under
local unitaries. We then compare the resulting optimal
$d\times d$ entanglement cloner
to a ``local'' cloning transformation that can be achieved by applying a
separate universal cloning machine to each component of the bipartite system.

\section{Covariant cloner under local unitaries in dimension $d\times d$}

Following the ideas presented in \cite{LNFC04}, we seek for a cloning
transformation that (i) preserves separability, and (ii) maximizes
the entanglement of the two clones resulting from any ME input state.
We will characterize a cloner by considering the transformation
of an input that is maximally entangled with a reference system. 
By projecting the reference onto
(the complex conjugate of) the input state, one gets the operation of the
cloner on this state.
The general form for such a cloning transformation is defined
in the computational basis $\{|i\rangle\}$ by the state 
\beq\label{cloner}
  |{\mathcal S}\rangle_{{\mathcal R}, a,b,{\mathcal A}}
   = \sum_{i,j,k,l} s_{ijkl} \, |i\rangle_{\mathcal R}
   \, |j\rangle_a \, |k\rangle_b \, |l\rangle_{\mathcal A}
\eeq
where $\mathcal R$ denotes the reference system, $a$ and $b$ stand for 
the two clones, and $\mathcal A$ corresponds to the ancilla.
All the summations here are $d^2$-dimensional since
all the states involved are $d^2$-dimensional bipartite states, e.g., $|i\rangle=|i_A\rangle|i_B\rangle$.
Thus, each index $i$, $j$, $k$, or $l$ 
actually represents a couple of indices
running each from 0 to $d-1$, e.g., 
$i=\{i_A,i_B\}$, with $i_A,i_B \in [0;d-1]$. Of course, the
index $A$ stands for Alice's component of the bipartite states, 
while $B$ stands for Bob's component.

As mentioned above, the joint state of the two clones and the ancilla
is obtained by performing an appropriate projection 
on the reference system.  Thus,
for an input state $|\Phi\rangle=\sum_in_i|i\rangle$,
the result of the cloning transformation is of the form
\beq \label{out}
  |\chi\rangle = \,
  _{\mathcal{R}}\langle\Phi^*|{\mathcal S}\rangle_{\mathcal R,a,b,{\mathcal A}}
    = \sum_{i,j,k,l} s_{ijkl}\,n_{i} \, |j\rangle_a \, |k\rangle_b \, |l\rangle_{\mathcal A} .
\eeq
Then, the state of any one of the clones is further obtained by tracing out
the ancilla and the other clone. This entanglement cloning transformation
is shown in Fig.~1.

\begin{widetext}

\begin{figure}[h]
\begin{center}
\includegraphics[width=0.9\textwidth]{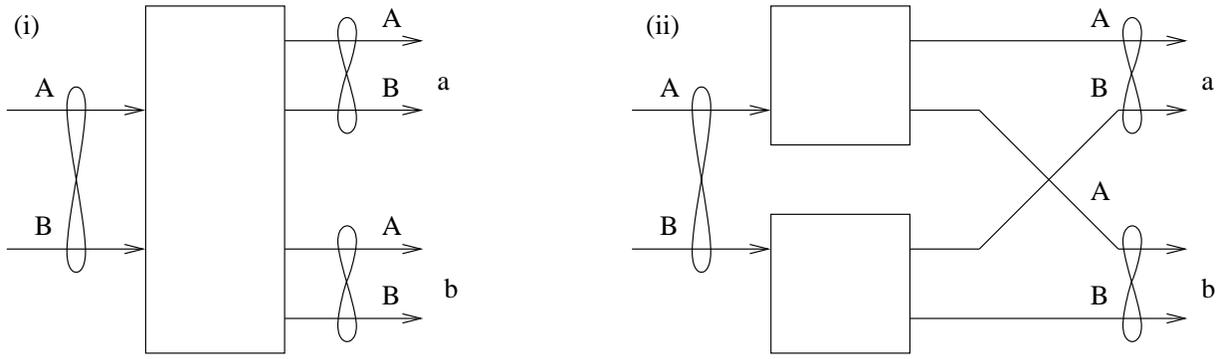}
\caption{(i) Optimal $d\times d$ entanglement cloner; 
(ii) ``local'' entanglement cloner, as defined in Sec.~IV.Here, A and B stand for
Alice's and Bob's part of the bipartite state, a and b stand for
two clones. Entanglement is indicated by double loops.}
\end{center}
\end{figure}

\end{widetext}

Next, we impose the following covariance condition on our cloning machine.
Since we know that any local unitary operation acting on the A and B 
components of a bipartite state preserves its entanglement,
we require that any such transformation acts similarly on the clones. This condition amounts to imposing
\beq \label{covcon}
  |{\mathcal S}\rangle_{{\mathcal R}, a,b,{\mathcal A}}
   = U^*\otimes U\otimes U\otimes U^*|\mathcal S\rangle_{\mathcal R,a,b,{\mathcal A}}
\eeq
where $U$ is the product of any two unitary transformations acting separately on each $d$-dimensional component
of the bipartite state, that is
\beq\label{U}
 U=U_A\otimes U_B
\eeq
where the indices $A$ and $B$ denote Alice's and Bob's components.
Defined in this way, the operator $U$ possesses a 
$SU(d) \otimes SU(d)$ symmetry.
The covariance property implies that $s_{ijkl}$ in (\ref{cloner}) 
is an invariant tensor of rank 4, which satisfies
\beq \label{tensors}
   s_{ijkl}=U^*_{ii'}U_{jj'}U_{kk'}U^*_{ll'}s_{i'j'k'l'}
\eeq
where $U^*$ denotes the matrix element-wise complex conjugate of $U$ with respect to the computational basis.
(Here, the summation over all repeated indices is implicit.)
The matrix elements $U_{ij}=\langle i|U|j\rangle$ satisfy $U^*_{ik}U_{kj}=\delta_{ij}$. 
If the transformation is real (i.e. $U_{ij}=U^*_{ij}$),
then the invariant tensors for Alice's component are of the general form
$\delta_{i_{A}j_{A}}\delta_{k_{A}l_{A}}$, or $\delta_{i_{A}l_{A}}\delta_{j_{A}k_{A}}$, or $\delta_{i_{A}k_{A}}\delta_{j_{A}l_{A}}$ (see \cite{ham89}).
For a complex transformation on Alice's component, the terms of the form $\delta_{i_{A}l_{A}}\delta_{j_{A}k_{A}}$
are excluded, so we are left with the two remaining terms.
Then, for a tensor product of (complex) unitary transformations 
on Alice's and Bob's components,
a general form for the invariant tensor is
\beqa \label{u1}
  \lefteqn{ s_{ijkl}} \nonumber\\
  & = & A \, \delta_{i_{A}j_{A}}\delta_{k_{A}l_{A}}\delta_{i_{B}j_{B}}\delta_{k_{B}l_{B}}
      + B \, \delta_{i_{A}j_{A}}\delta_{k_{A}l_{A}}\delta_{i_{B}k_{B}}\delta_{j_{B}l_{B}} \nonumber\\
  & + & C \, \delta_{i_{A}k_{A}}\delta_{j_{A}l_{A}}\delta_{i_{B}k_{B}}\delta_{j_{B}l_{B}}
      + D \, \delta_{i_{A}k_{A}}\delta_{j_{A}l_{A}}\delta_{i_{B}j_{B}}\delta_{k_{B}l_{B}}.
\nonumber\\
\eeqa
For a symmetric cloner, the output state must be invariant under the
interchange of the two clones, i.e., under permutations $(j_{A},j_{B}) \leftrightarrow (k_{A},k_{B})$. This implies that
$A=C$ and $B=D$, so we are left with only two parameters $A$ and $B$ to be determined.

\section{Optimal entanglement cloner in dimension $d\times d$}

The covariance condition (\ref{covcon})
guarantees that our QCM transforms all states 
which are equivalent up to local unitaries 
(which have therefore the same entanglement)
into equally entangled clones.
In particular, the clones of all ME states will be equally entangled.
Then, a cloner that is optimized on a particular ME input state 
will be optimal for all ME states. We choose as an initial $d\times d$
ME state
\beq \label{MEin}
  |\Phi\rangle=\sum_{i_A,i_B=0}^{d-1}n_{i_A i_B}|i_A\rangle |i_B\rangle 
\eeq
where $n_{i_A,i_B}=\delta_{i_Ai_B}/\sqrt{d}$. 
As we shall show later on, we can maximize the entanglement of the clones
simply by optimizing our QCM in terms of the fidelity of the clones, 
\beq \label{F}
  F = \langle\Phi|\rho_a|\Phi\rangle \, ,
\eeq
where
\beq \label{clona}
  \rho_a = {\rm Tr}_{{\mathcal A},b}\left[ |\chi \rangle \langle \chi | \right]
\eeq
is the state of clone $a$.
For the ME state (\ref{MEin}), this fidelity is found to be
\beq \label{F1}
  F = |A|^2(d^2+3)+4|B|^2+4\, \Re (AB^*) \frac{d^2+1}{d}.
\eeq
Taking into account the normalization condition
for the joint output state $|\chi\rangle$,
\beq \label{norm}
  2\left(|A|^2+|B|^2\right)(d^2+1)+8d\, \Re (AB^*) = 1 \, ,
\eeq  
we can maximize the fidelity, Eq. (\ref{F1}), which yields
\beq \label{Fmax}
  F = \frac{1}{4}
          \left(\frac{d^2+1}{d^2-1}
               +\sqrt{1+\frac{4}{d^2}
                        \left(\frac{d^2-2}{d^2-1}\right)^2
                      }
           \right)  .
\eeq
Note that, for $d=2$, this result coincides with the maximal fidelity
of the entanglement cloner for two qubits obtained in \cite{LNFC04}, namely
\beq \label{fidel2}
  F=\frac{5+\sqrt{13}}{12}\approx 0.7171.
\eeq
The corresponding solution in $d$ dimensions is
\beqa \label{A}
  A & = & \frac{d\sqrt{1+Y(d)}-\sqrt{1-Y(d)}}{2(d^2-1)},\\
  B & = & -\frac{d\sqrt{1-Y(d)}+\sqrt{1+Y(d)}}{2(d^2-1)},\label{B}
\eeqa
where
\beq
  Y(d)=\left(1-\frac{(d^2-2)^2}{d^2(d^2-1)^2+4(d^2-2)^2}\right)^{1/2} .
\eeq

\section{Comparison with other cloners}

We compare the fidelity achieved by our optimal $d\times d$-dimensional entanglement cloner, Eq.~(\ref{Fmax}), with that of the 
universal cloner \cite{C00}
\beq \label{Fu}
  F_u= \frac{1}{2}+\frac{1}{d^2+1},
\eeq
as well as that of the optimal real cloner  \cite{NC03} 
\beq \label{Fr}
  F_r= \frac{1}{2}+\frac{\sqrt{d^4+4d^2+20}-d^2+2}{4(d^2+2)}.
\eeq
In order to make this comparison consistent,
we have obtained formulae (\ref{Fu}) and (\ref{Fr}) 
by replacing the argument $d$ by $d^2$ in the original formulae.
This is done because, in our consideration, 
the dimension $d$ stands for the dimension of each component
of the bipartite input state, so that
the total dimension of our input state is $d^2$.

In Table~I, we compare the fidelity $F$ of our entanglement cloner 
with $F_u$ and $F_r$, for several values of the dimension $d$.
Our cloner performs better than the universal cloner
in $d^2$ dimensions for all $d$, which is obviously 
due to the fact that the ME states
span only a subset of the entire set of $d^2$-dimensional states. In contrast,
the real $d^2$-dimensional cloners outperform our cloners, except if $d=2$ where they coincide \cite{LNFC04}. This can be interpreted by noting that the
set of $d^2$-dimensional real states is generated by $SO(d^2)$, with 
$(d^2(d^2+1)/2)-1$ real degrees of freedom, while the set of ME states is generated by $SU(d)\times SU(d)$, with $(d^2-1)^2$ real degrees of freedom.
For $d=2$, they coincide, so that our cloner provides 
the same fidelity as that of the real cloner in dimension 4, namely Eq.~(\ref{fidel2}). This is
related to the fact that the set of ME 2-qubit states is isomorphic 
to the set of 4-dimensional real states \cite{LNFC04}.
For $d>2$, the set of ME states is in some sense ``larger'' than
the set of real states for $d>2$, so that the achievable fidelity
of the entanglement cloner is lower.  
The fidelity of our cloner drops faster than that of the real cloner with increasing $d$, but remains always higher than
the fidelity of the universal cloner. As expected, in the limit $d\rightarrow\infty$,
all three fidelities tend to the asymptotic value 1/2. 
In this limit, all quantum cloners can be interpreted simply as a classical
transformation that maps the original state to one of the clones, 
chosen at random, the other clone being prepared in a maximally mixed state.

\begin{center}
\begin{table}
\begin{tabular}{|c|c|c|c|c|}
\hline
$\, d\times d\, $ & $F_r$ & $F$ & $F_u$ & $F_{\rm loc}$\\
\hline
2 $\times$ 2& \,0.7171 \, & \, 0.7171 \, &\, 0.7000 \, & \, 0.5833\, \\
3 $\times$ 3& 0.6069 & 0.6019 & 0.6000 & 0.4583\\
4 $\times$ 4& 0.5617 & 0.5592 & 0.5588 & 0.4000\\
5 $\times$ 5& 0.5385 & 0.5386 & 0.5385 & 0.3667\\
6 $\times$ 6& 0.5277 & 0.5271 & 0.5270 & 0.3452\\
\hline
\end{tabular} 
\caption{Optimal fidelity $F$ of the $d\times d$ entanglement cloner
for various dimensions $d$. It is compared to the fidelity of the real cloner
$F_r$ and universal cloner $F_u$, both in $d^2$ dimensions, and to the fidelity of the $d\times d$ ``local'' cloner $F_{\rm loc}$.
The fidelities are shown in decreasing order. }
\end{table}
\end{center}

\vspace{-1cm}
Interestingly, we may also compare Eq.~(\ref{Fmax}) to the fidelity of
a ``local cloner'' obtained by applying a cloner separately
to Alice's and Bob's components (see Fig. 1). 
Since the state of Alice's or Bob's
subsystem is maximally mixed (hence non-polarized) when the bipartite state
is ME, it is natural to use a universal $d$-dimensional cloner. We may observe
that if we consider a cloning transformation that performs such a local universal cloning, then it is represented by a joint state
of the same type as Eq. (\ref{cloner}), see \cite{C00}. 
The only difference is that 
in the expression (\ref{u1}) for the tensor $s_{ijkl}$,
all coefficients must be equal, i.e., $A=B=C=D$.
The normalization condition (\ref{norm}) then gives immediately 
$A=1/(2(d+1))$. Substituting this expression into Eq. (\ref{F1})  
results in the fidelity 
\beq \label{Ful}
  F_{\rm loc}
    = \frac{1}{4}+\frac{d+2}{2d(d+1)}
\eeq
for the local cloner. 
This fidelity is compared in Table~1 with that of the other cloners.
It appears that cloning Alice's and Bob's parts locally
leads to a much lower fidelity. Note that for $d=2$, the value of the fidelity 
$F_{\rm loc}$ in Table 1 coincides with the value $7/12$ obtained in \cite{LNFC04}. 
In the limit $d\rightarrow\infty$, this fidelity tends to 1/4, 
which can be easily understood in the classical picture above. To contribute
to the fidelity, both cloners indeed need to map Alice's and Bob's components 
of the original state onto the right clone, which only happens 
with probability $(1/2)^2=1/4$.

\section{Entanglement of formation of the clones}

In order to investigate the entanglement properties of our cloning transformation, we will use as an entanglement measure for the clones
the entanglement of formation \cite{BDSW96}, which was computed for
several classes of states that are invariant under some groups of local symmetries \cite{VW01}. In particular, we will be interested in
the class of states that are invariant 
under the transformations $U\times U^*$ for all $U$, called {\it isotropic} states in \cite{HH99,VW01}. These states
may be written in a general form as \cite{TV00}
\beq\label{WS}
  \rho = \frac{1-F}{d^2-1}({\openone}-|\Phi\ra\la\Phi|)+F|\Phi\ra\la\Phi|
\eeq
where ${\openone}$ is the identity and $|\Phi\ra$ is given by Eq.~(\ref{MEin}).
Due to the covariance condition (\ref{covcon}), our QCM transforms 
$U\times U^*$ invariant states into into states that are also invariant under 
$U\times U^*$. We can check that, by cloning the particular ME state $|\Phi\rangle$, which is $U\times U^*$ invariant, we obtain a clone of the form
\beqa
  \rho_a
  & = & 
        \Big((d^2+2)|A|^2+2|B|^2+4d\, \Re(A^*B)\Big) \,
	|\Phi\rangle \langle\Phi| \nonumber \\ 
\label{rhoa}
  & + & \Big(|A|^2+2|B|^2+\frac{4}{d}\, \Re(A^*B)\Big) \, \openone
\eeqa
which is indeed an isotropic state and is consistent with Eq.~(\ref{F1}).
Hence, as a consequence of our covariance condition, all ME states, which 
can be obtained from $|\Phi\rangle$ by applying local unitaries,
are cloned onto isotropic states.
For this class of states, the entanglement of formation is known 
for arbitrary dimensions \cite{VW01,TV00}
\beq \label{EF}
  E_F(\rho)=\left\{\begin{array}{l l }
                        0, & F \le \frac{1}{d} \\ [2ex]
                        R_{1,d-1}(F), & F\in[\frac{1}{d},\frac{4(d-1)}{d^2}] \\ [2ex]
                        \frac{d\log_2(d-1)}{d-2}(F-1)+\log_2 d, & F\in[\frac{4(d-1)}{d^2},1]
                   \end{array}
            \right.
\eeq
where 
\beqa \label{R}
  R_{1,d-1}(F)
  & = & H_2(\gamma(F))+[1-\gamma(F)]\log_2(d-1) , \\ [2ex]
\label{g}
  \gamma(F)
  & = & \frac{1}{d}\left[\sqrt{F}+\sqrt{(d-1)(1-F)}\right]^2  , \\ [2ex]
\label{H2}
  H_2(p)
  & = & -p\log_2(p)-(1-p)\log_2(1-p) .
\eeqa
As shown in Fig.~2, the entanglement of formation $E_F(\rho)$
is a monotonically increasing function of the fidelity $F$
for isotropic states in any dimension $d$.
Therefore, by optimizing our QCM in terms of fidelity we maximize,
at the same time, the entanglement of the clones 
measured by their entanglement of formation.
The circles in Fig.~2 correspond to the maximal fidelity $F$ 
that is achieved by our entanglement cloner, Eq.~(\ref{Fmax}).
They show as well the corresponding entanglement of formation
in this dimension. 
The crosses mark the crossover between the expression of the fidelity 
corresponding to the second and third lines of Eq.~(\ref{EF}).
One notes that only for $d\ge 7$ there are values of the fidelity 
for which the entanglement of formation 
has to be evaluated according to the third line of Eq.~(\ref{EF}).

\begin{figure}[h]
\begin{center}
\includegraphics[width=0.45\textwidth]{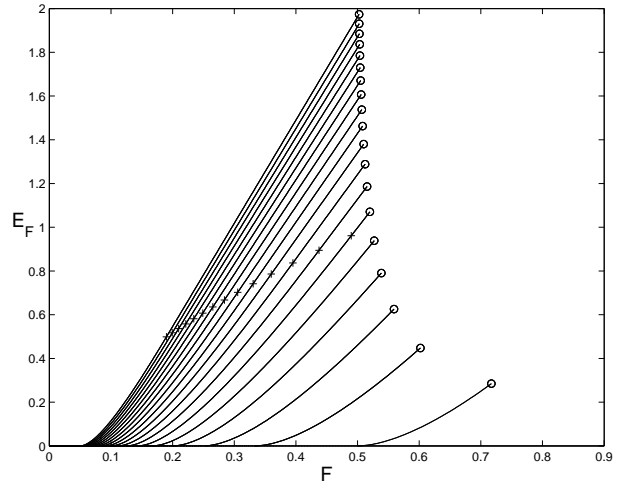}
\caption{Entanglement of formation $E_F$ of the clone of a maximally-entangled input state versus the fidelity $F$ of the clone
for various dimensions $d=2-20$ (the lowest curve corresponds to $d=2$, while
the highest curve corresponds to $d=20$). The circles show the maximum achievable fidelity and the corresponding entanglement of formation.
The crosses mark the crossover between the expression of the fidelity
corresponding to the second and third lines of Eq.~(\ref{EF}).}
\end{center}
\end{figure}

In order to visualize how the entanglement itself is cloned,
we plot in Fig.~3 the entanglement of formation $E_F$ of the clones
as a function
of the entanglement of formation of the input ME state $E_{\rm in}$, which is 
simply the von Neumann entropy of the reduced density matrix $E_{in}=\log_2 d$.
We note that the entanglement of the clones is always less than one half the entanglement of the input state, while it asymptotically 
approaches this value for large $d$. 
The apparent ``discontinuity'' (if one can say so for
a discrete graph) corresponds to $d=7$, i.e., the crossover between the second and the third line of Eq. (\ref{EF}) when calculating the entropy of formation. In the limit of $d\to\infty$, the third line of Eq. (\ref{EF})
tends to $E_F=F\, \log_2 d=F\, E_{\rm in}$. Since the cloner can be viewed,
in this limit, as a classical random distribution process associated with
a fidelity $F=1/2$, then the entanglement of each clone tends to one half 
the entanglement of the initial state.

\begin{figure}[h]
\begin{center}
\includegraphics[width=0.45\textwidth]{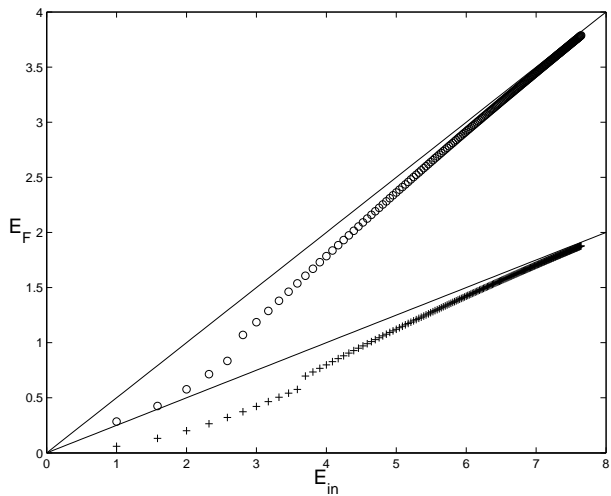}
\caption{Entanglement of formation $E_F$ of the clones 
of a maximally-entangled state obtained by the optimal (non-local) cloner ($\circ$) and the local cloner ($+$) versus the entanglement 
of the input state $E_{in}$ for various dimensions $d=2-200$. The apparent ``discontinuity'' in both curves is due
to the crossover from the second to the third line of Eq.~(\ref{EF}) 
for $d\ge7$ (optimal cloner) and $d\ge 13$ (local cloner).
Solid lines represent the asymptotics of $E_F$ for large $d$ in both cases.}
\end{center}
\end{figure}

In Fig.~3, we also plot the entanglement of formation resulting from
the ``local'' cloner discussed above. Recall that this cloner differs 
from our optimal (non-local) entanglement cloner only by setting $A=B$.
Therefore it is also covariant, satisfying Eq. (\ref{covcon}), and all our arguments about the entanglement of formation of the clones are applicable to this cloner as well. Thus, using the fidelity of the clones (\ref{Ful}),
we may plot the entanglement of formation of the clones. We see
that the local cloner leads to a lower entanglement of formation, and 
even the asymptotics of $E_F$ for large $d$ is no more than
one fourth the entanglement of the input state.
The reason is that in the limit of large $d$, the classical 
random distribution only succeeds with probability 1/2 
independently for Alice's and Bob's 
components, so the fidelity is 1/4. Hence, $E_F\to E_{\rm in} /4$.
These observations confirm that by increasing the dimensionality,
we make the system behavior look more and more classical.

\section{Separability conservation}

The last point to check is that our cloner does not create entanglement 
by itself, that is, it clones separable states into separable states.
First, an important observation is that our cloner is
such that the input-to-single-clone transformation is a PPT map. 
Using Eqs. (\ref{cloner}) and (\ref{u1}), and tracing
the joint state $|S\rangle_{{\mathcal R},a,b,{\mathcal A}}$
over the ancilla ${\mathcal A}$ and one of the clones, say $b$,
we arrive at the following expression for the state 
of the reference and the other clone 
\beqa
\lefteqn{  S_{{\mathcal R},a} 
= |A|^2 (\openone_A \otimes \openone_B)_{{\mathcal R},a} }
\hspace{0.5cm} \nonumber\\
&+&  d^2 \left((d^2+2)|A|^2+2|B|^2+4d\, \Re (A^*B)\right) \nonumber\\
&&\times (|\phi\rangle_A \langle\phi|_A \otimes |\phi\rangle_B \langle\phi|_B)_{{\mathcal R},a} \nonumber\\
&+& d \left(d|B|^2+2\, \Re(A^*B)\right) \nonumber\\
&&\times (|\phi\rangle_A \langle\phi|_A \otimes \openone_B + \openone_A \otimes |\phi\rangle_B \langle\phi|_B)_{{\mathcal R},a}
\eeqa
where $\openone_A$  is the identity operator 
in the joint space of Alice's component of the reference
${\mathcal R}$ and clone $a$, while $|\phi\rangle_A= d^{-1/2} \sum_{i=0}^{d-1}
|i\rangle_{A,{\mathcal R}} |i\rangle_{A,a}$ is a ME state in this
same space. (The same notations are used
for Bob's analog quantities $\openone_B$ and $|\phi\rangle_B$.)
The cloning map is thus PPT since $(S_{{\mathcal R},a})^{T_B}\ge 0$,
where $T_B$ stands for the partial transposition with respect 
to Bob's components of the reference ${\mathcal R}$ and clone $a$. 
This PPT property ensures that the cloning
of any isotropic state cannot increase its fidelity, hence its entanglement
of formation \cite{R01}. In particular, all separable isotropic states
are necessarily transformed into separable clones.

In order to generalize this separability conservation property 
to all separable input states, outside the restricted class of isotropic
states, we consider the cloning of a product state $\rho_A\otimes\rho_B$. 
By tracing $(\rho_A\otimes\rho_B)^T S_{{\mathcal R},a}$
over the reference ${\mathcal R}$ , we obtain for the first clone 
a state of the form
\beqa 
\lefteqn{ \rho_{a} = |A|^2 (\openone_A\otimes \openone_B)_a }  \label{clon0}
\nonumber\\
  & + & \left((d^2+2)|A|^2+2|B|^2+4d\, \Re (A^*B)\right) (\rho_A\otimes\rho_B)_a \nonumber \\
  & + & \left(d|B|^2+2\, \Re(A^*B)\right)
        \left(\rho_A\otimes \openone_B+\openone_A\otimes \rho_B\right)_a
\eeqa
where $\openone_A$ and $\openone_B$ are identities in Alice's and Bob's 
subspaces of clone $a$, respectively.
Since all terms in (\ref{clon0}) are product states 
and all coefficients are positive semi-definite for all $d$, we verify
that $\rho_a$ is separable. By linearity of the trace, 
this result also holds for any linear combination 
$\sum_i p_i\rho^A_i\otimes\rho^B_i$ with $p_i\ge 0$ and $\sum_i p_i=1$), that is for a generic separable
state. Thus, we can conclude that our entanglement cloner 
transforms all initially separable states into separable clones.

\section{Conclusion}

In conclusion, we have constructed an optimal (symmetric) entanglement cloner, 
which is universal over the set of $d \times d$ ME states 
in arbitrary dimension $d$. On one hand, all separable 
input states are cloned by this cloner into separable states.
In addition, the entanglement of isotropic states cannot be increased
by the cloner (and we conjecture this property holds in general for any
input state). On the other hand, the entanglement of the clones 
of ME input states is maximum. The optimization of the parameters 
of our QCM was performed by maximizing the fidelity of the clones, 
but the monotonic behavior of the entanglement of formation 
as a function of the fidelity for isotropic states guarantees 
that such an optimization maximizes the entanglement of the clones
at the same time. We expect that entanglement is cloned ``monotonically''
for all (including non-isotropic) states, 
that is,  higher entangled states result in higher entangled clones,
and therefore the ME input states are those
which generate the clones with the maximum achievable entanglement.
If this very natural assumption is right, then, based on our result, 
one can state that the maximal entanglement attainable by cloning
is always below one half of the entanglement of the input state
and saturates this value in the limit of large dimension $d$.
This is consistent with the idea that, since our QCM transforms 
separable states into separable clones, no additional entanglement 
is produced by cloning, so we can only split the entanglement 
of the input state between the two clones. 
This explains as well the asymptotic value of one fourth the initial entanglement for the local cloner at the limit of large $d$. 
It is natural to expect all these conclusions
to be valid for asymmetric entanglement cloners as well.

\acknowledgments
We thank Louis-Philippe Lamoureux and Jaromir Fiurasek
for helpful discussions. We acknowledge the support
from the European Union under the projects RESQ (Grant No.~IST-2001-37559)
and SECOCQ (Grant No.~IST-2002-506813), from the Communaut\' e Fran\c caise de la Belgique
under the ``Action de Recherche Concert\' ee'' nr. 00/05-251, and from
IAP program of Belgian federal government under Grant No. V-18.

\end {document}